\documentclass[twocolumn,showkeys,showpacs]{revtex4-1}
\usepackage[cp1251]{inputenc}
\usepackage{amsmath}
\usepackage{amsfonts}
\usepackage{amssymb}
\usepackage{graphicx}
\usepackage{textcomp}
\usepackage{color}
\begin{document}

\title{Fitting of the TB-SMA interatomic potentials for Pt/Cu(111) surface alloy}
\author{S.A. Dokukin}
\author{S.V. Kolesnikov}
\email{kolesnikov@physics.msu.ru}
\author{A.M. Saletsky}
\author{A.L. Klavsyuk}

\affiliation{Faculty of Physics, Lomonosov Moscow State University, Moscow 119991, Russian Federation}

\begin{abstract}	
In this paper we present new parameters of the TB-SMA interatomic potentials for the Pt/Cu(111) surface alloy. The parameters are fitted using both the experimental and {\it ab initio} data. The potentials reproduce not only the bulk properties of copper and platinum, but also the energy characteristics of the Pt/Cu(111) surface alloy. The potentials can be used for the simulations of the growth of the Pt/Cu(111) surface alloy on the atomic scale.
\end{abstract}	



\date{\today}
	
\maketitle

Theoretical methods for studying the formation of surface alloys can be divided into several groups. Most precise and reliable are the {\it ab initio} methods. They are often based on the density functional theory~\cite{Kohn1996}. Unfortunately, such methods are too complicated. They can only be used for the calculation of the typical energies or to describe several initial steps of the system's evolution. Another approach is using of the semiempirical potentials for the investigation of the atomic properties of alloys on the long-time scale. For instance, the Bozzolo-Ferrante-Smith (BFS) potentials were used to describe the Cu$_3$Pt alloy formation during the annealing of the Cu(100) substrate covered with the Pt monolayer~\cite{Demarco2003}. The BFS potentials are based on the values of the experimental and {\it ab initio} parameters of pure materials and, therefore, can describe some of their properties. However, they have only one fitting parameter for interaction of different types of atoms. Thus, it is almost impossible to fit this parameter in order to describe all the properties of the surface alloy. The embedded-atom method (EAM) and modified embedded-atom method (MEAM) potentials were also proposed for this system~\cite{Luyten2007,Mu2017}. Such potentials are quite popular because they can precisely reproduce the typical energies and the bulk characteristics of the pure materials~\cite{Baskes1992}. They can also be used to evaluate the parameters of bulk or dilute alloys~\cite{Mo2016,Zhou2017}. Unfortunately, the available potentials~\cite{Luyten2007,Mu2017} for the Pt/Cu system neglect the bulk properties of pure Cu and Pt. Incorrect values of the lattice constant and elastic constants of the substrate will lead to an additional energy of elastic deformation. Therefore, an accurate calculation of the adsorption energies and diffusion barriers of Pt atoms on a Cu substrate is unlikely. Another type of interatomic potentials are based on the tight binding method in the second moment approximation (TB-SMA)~\cite{Cleri1993}. This approximation was used to simulate the formation of various heterogeneous alloys: Co/Cu~\cite{Kolesnikov2009,Stepanyuk2008}, Fe/Cu~\cite{Negulyaev2008_2}, Ni/Cu~\cite{Meyerheim2008}, Cu/Ti~\cite{Qin2010}, Co/Pt~\cite{Goyhenex2000}, Co/Au~\cite{Goyhenex2001}, Co/Ag~\cite{Guo2005}, Ag-Au-Pd-Pt~\cite{Wu2016}. The TB-SMA potentials can simultaneously reproduce the atomic adsorption energies and the diffusion barriers, and precisely describe the bulk structure of substrates. Therefore, it is possible to simulate such processes as formation of nanostructures~\cite{Kolesnikov2009}, vibration of clusters~\cite{Borisova2008}, mesoscopic relaxations~\cite{Stepanyuk2000}, phase stability and segregation~\cite{Zhao2017}. It was also shown that the TB-SMA potentials are more suitable for the simulation of the nanocontacts than the EAM~\cite{Pu2007,Klavsyuk2015}.

We use the interatomic TB-SMA potentials~\cite{Rosato1989,Cleri1993,Levanov2000}. In this approximation, the attractive term $E_b^i$ (band energy) contains the many-body interaction. The repulsive part $E_r^i$ is described by pair interactions (Born-Mayer form). The cohesive energy $E_C$ is the sum of the band energy and the repulsive part:
\begin{equation}
\label{eqEc}
E_C=\sum\limits_{i} (E_b^i+E_r^i),
\end{equation}
\begin{equation}
\label{eqEb}
E_b^i=-\sqrt{\sum_j\xi_{\alpha\beta}^2\exp\left[-2q_{\alpha\beta}\left(\frac{r_{ij}}{r_0^{\alpha\beta}}
-1\right)\right]f_c(r_{ij})},
\end{equation}
\begin{multline}
\label{eqEr}
E_r^i=\sum_j\left[A_{\alpha\beta}^1\left(\frac{r_{ij}}{r_0^{\alpha\beta}}-1\right)
+A_{\alpha\beta}^0\right]\\
\exp\left[-p_{\alpha\beta}\left(\frac{r_{ij}}{r_0^{\alpha\beta}}-1\right)\right]f_c(r_{ij}),
\end{multline}
where $r_{ij}$ is the distance between the atoms $i$ and $j$; $\alpha$ and $\beta$ are types of the atoms; $\xi_{\alpha\beta}$ is an effective hopping integral; $p_{\alpha\beta}$ and $q_{\alpha\beta}$ describe the decay of the interaction strength with distance between the atoms; and $r_0^{\alpha\beta}$, $A_{\alpha\beta}^0$ and $A_{\alpha\beta}^1$ are adjustable parameters of an interatomic interaction. The cut-off function $f_c(r_{ij})$ has the form~\cite{Brooks1983}: $f_c(r_{ij})=0$ if $r_{ij}\leq R_{on}$, $f_c(r_{ij})=1$ if $r_{ij}\geq R_{off}$, and
\begin{equation}
f_c(r_{ij})=\frac{(R_{off}^2-r_{ij}^2)^2(R_{off}^2-3R_{on}^2+2r_{ij}^2)}{(R_{off}^2-R_{on}^2)^3}
\end{equation}
if $R_{on}<r_{ij}<R_{off}$, where $R_{off}=6.5$~\AA~and $R_{on}=6.0$~\AA~are cut-off distances.

\begin{table*}[!t]
\caption{Data used for the fitting of the TB-SMA potentials, values calculated with the optimized potentials and values calculated with the potentials from~\cite{Garbouj2010}. Lattice constant $a$, cohesive energy $E_c$ and bulk modulus $B$ are taken form~\cite{Cleri1993}. The binding energies and the diffusion barriers are calculated with the VASP code~\cite{Kresse1993}. The atomic configurations used in the fitting procedure are shown in Fig.~\ref{met_fig1}.}  \label{table1}
\begin{center}
\begin{tabular}{ccccccc}
\hline
\hline
\rule{0cm}{4.5mm}
~&~Quantity~&~Configuration        ~&~~~~Data~~~~&~~~~Fitted~~~~&~~~~Values~~~~             \\
~&~        ~&~(Fig.~\ref{met_fig1})~&~     ~&~~~~value~~~~&~~~~from~\cite{Garbouj2010}~~~~\\
\hline
\rule{0cm}{4.5mm}
Pt                 & $a$    &    & 3.924 \AA   & 3.925 \AA  & 3.930 \AA\\
\rule{0cm}{4.5mm}
(fcc)       & $E_c$   &   & 5.853 eV    & 5.853 eV & 5.871 eV\\
\rule{0cm}{4.5mm}
~                  & $B$   &     & 2.88 Mbar   & 2.89 Mbar & 2.86 Mbar\\
\hline
\rule{0cm}{4.5mm}
Cu-Pt              & $E_{1on}$   &  1   & -4.958 eV & -4.835 eV & -3.950 eV\\
\rule{0cm}{4.5mm}
~                  & $E_{2on1}$  &  2   & -0.335 eV & -0.454 eV & -0.942 eV\\
\rule{0cm}{4.5mm}
~                  & $E_{1in}$    & 3   & -2.963 eV & -2.939 eV & -2.054 eV\\
\rule{0cm}{4.5mm}
~                  & $E_{2in1}$   & 4   &  0.154 eV & 0.125  eV & -0.010 eV\\
\rule{0cm}{4.5mm}
~                  & $E_{2in3}$   & 5   &  0.019 eV & 0.055  eV & 0.013 eV\\
\rule{0cm}{4.5mm}
~                  & $E_{1on}^{st}$ & 6 & -5.899 eV & -5.900 eV & -4.730 eV\\
\rule{0cm}{4.5mm}
~                  & $E_{2on1}^{st}$ & 7 & -0.277 eV & -0.117 eV & -0.554 eV\\
\rule{0cm}{4.5mm}
~                  & $E_{1in}^{st}$ & 8 & -2.666 eV & -3.024 eV & -1.770 eV\\
\rule{0cm}{4.5mm}
~                  & $E_{2in1}^{st}$ & 9 &  0.152 eV & 0.352  eV & 0.001 eV\\
\rule{0cm}{4.5mm}
~                  & $E_{2in3}^{st}$ & 10 &  0.015 eV &  0.132 eV & -0.002 eV\\
\rule{0cm}{4.5mm}
~                  & $E_d^{step}$   &  &  0.470 eV &  0.457 eV & 0.252 eV\\
\rule{0cm}{4.5mm}
~                  & $E_d^{surf}$  &   &  1.100 eV &  1.060  eV & 0.630 eV\\
\hline
\hline
\end{tabular}
\end{center}
\end{table*}

Parameters for Cu-Cu interaction were taken from Ref.~\cite{Negulyaev2008_1}. The reliability of the Cu-Cu potentials for different atomic structures (supported islands, surface and bulk vacancies, nanocontacts) has been demonstrated~\cite{Lysenko2002,Kolesnikov2009_2,Siahaan2016,Stepanyuk2004}.

\begin{figure}[!th]
\begin{center}
\includegraphics[width=0.95\linewidth]{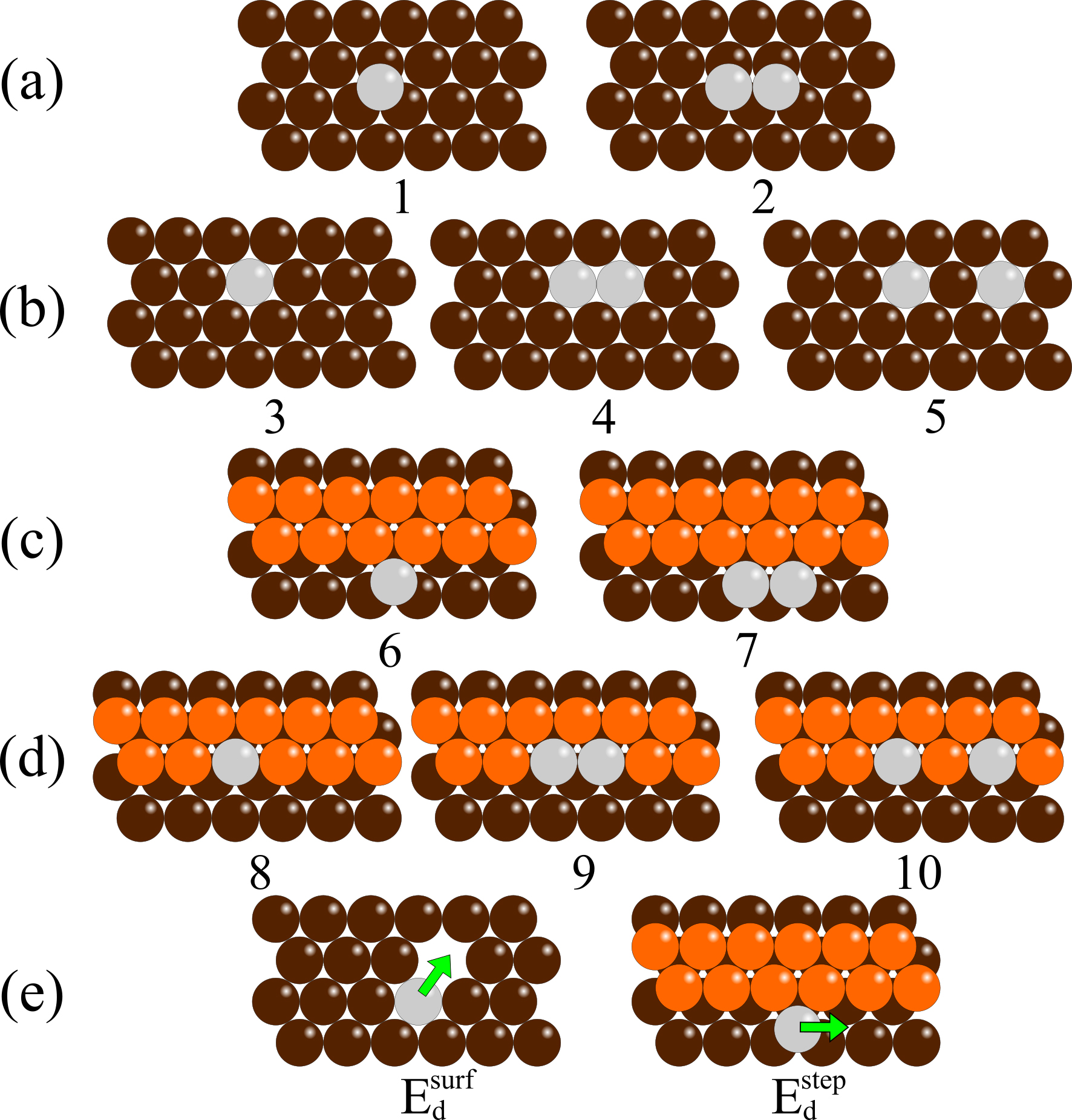}
\caption{\label{met_fig1} Schematic view of the topmost layers of the calculation cells used to fit the Cu-Pt potentials: (a) Pt adatoms on the surface, (b) Pt adatoms in the surface, (c) Pt adatoms near Cu step edge, (d) Pt adatoms in the Cu step edge, (e) the barriers used in the fitting procedure. Orange and brown balls symbolize Cu atoms and gray balls - Pt atoms.}
\end{center}
\end{figure}

The Pt-Pt and Cu-Pt parameters are optimized simultaneously by including in the fit the experimental data for the bulk properties of Pt (lattice constant $a$, cohesive energy $E_c$, and bulk modulus $B$)~\cite{Cleri1993} and the results of the first-principles DFT calculations. The following energies were included in the database (see Fig.~\ref{met_fig1} and Table~\ref{table1}): the binding energies of adatom $E_{1on}$ and dimer $E_{2on1}$ on the Cu(111) surface, the binding energies of adatom $E_{1in}$ and dimers on the first $E_{2in1}$ and third $E_{2in3}$ neighbors in the topmost layer of the Cu(111) surface and the similar energies in the case of the stepped Cu(111) surface. The binding energies of adatoms were calculated from the formula
\begin{equation}
\label{binding_energy1}
E = E_1 - E_0,
\end{equation}
where $E_1$ is the total energy of the calculation cell with the Pt adatom and $E_0$ is the total energy of the substrate without the adatom. To calculate binding energies of the dimers we used the following formula
\begin{equation}
\label{binding_energy2}
E = E_2 - 2E_1 + E_0,
\end{equation}
where $E_2$ is the total energy of the calculation cell with two Pt adatoms. We also include in our database the following two barriers: the barrier for the jump of Pt adatom embedded in the topmost layer of the Cu(111) surface $E_d^{surf}$, and the barrier for the jump of the Pt adatom along the Cu step edge $E_d^{step}$ (see Fig.~\ref{met_fig1}).

\begin{table}[b]
\caption{Parameters of interatomic interactions.}  \label{table2}
\begin{center}
\begin{tabular}{cccc}
\hline
\hline
~~~Parameter~~~&~~Cu-Cu\cite{Negulyaev2008_1}~~&~~Cu-Pt~~&~~Pt-Pt~~~~\\
\hline
\rule{0cm}{4.5mm}
$A^1$ (eV) &   0.000   &   0.049   &  -0.422  \\
$A^0$ (eV) &   0.085   &   0.241   &   0.120  \\
$\xi$ (eV) &   1.224   &   2.093   &   1.995   \\
$p$        &   10.939  &   9.958   &   9.560  \\
$q$        &   2.280   &   3.702   &   4.010   \\
$r_0$ (\AA)&   2.556   &   2.560   &   2.929   \\
\hline
\hline
\end{tabular}
\end{center}
\end{table}

All the DFT calculations were performed with the Vienna {\it ab initio} simulation package (VASP)~\cite{Kresse1993} with the generalized gradient approximation (GGA)~\cite{Wang1991,Perdew1992}. The substrate has been modeled as the periodically repeated slabs consisting of the six atomic layers separated by a sufficiently thick vacuum space of 16~\AA. Each atomic layer consisted of the 6$\times$4 atoms. Resulting surface of the slab was Cu(111). The positions of the atoms in the four topmost layers of the substrate were optimized using scalar-relativistic calculations until the forces on all unconstrained atoms were converged to less than 0.01~eV/\AA. The same structure was used as a substrate for the energies calculations with the TB-SMA potentials. Its relaxation was carried out by the molecular statics method. To calculate the barriers for the fitting procedure we used the same computational cell as for the calculations of the energies.

The set of data used to define the parameters of the interatomic TB-SMA potentials and the corresponding values calculated by means of the optimized potential are given in Table~\ref{table1}. The bulk properties, surface properties, and diffusion barriers are well reproduced. We also calculated the same values using the parameters taken form~\cite{Garbouj2010}: Cu-Cu and Pt-Pt potentials were fitted to the bulk properties of Cu and Pt, respectively, and parameters of the Cu-Pt potential were taken as averaged parameters. In the last case the surface properties and the barriers are reproduced worse. We see that it is critically important to include the surface properties in the fitting procedure. The fitted parameters of interatomic interactions are presented in Table~\ref{table2}.

In summary, we have fitted parameters of the interatomic TB-SMA potentials for the Pt-Cu system. These potentials reproduce bulk properties of copper and platinum, and energy characteristics of Pt/Cu(111) surface alloy. We believe that our parameters of the interatomic TB-SMA potentials will be useful for solution of many problems associated with the evolution and the dynamical properties of the Pt-Cu systems.

\section*{Acknowledgements}
The research is carried out using the equipment of the shared research facilities of HPC computing resources at Lomonosov Moscow State University~\cite{NIVC}.

\end{document}